\documentclass[conference,a4paper]{IEEEtran}
\usepackage{setspace}
\usepackage{amsmath,amsfonts}
\usepackage{algorithmic}
\usepackage{array}
\usepackage[caption=false,font=normalsize,labelfont=sf,textfont=sf]{subfig}
\usepackage{textcomp}
\usepackage{stfloats}
\usepackage{url}
\usepackage{verbatim}
\usepackage{graphicx}
\def\BibTeX{{\rm B\kern-.05em{\sc i\kern-.025em b}\kern-.08em
    T\kern-.1667em\lower.7ex\hbox{E}\kern-.125emX}}
\usepackage{balance}
\usepackage[noadjust]{cite}
\usepackage{moreverb}
\usepackage{widetext}
\usepackage{lipsum}
\usepackage{lingmacros}
\usepackage{tree-dvips}
\usepackage{varioref}
 \usepackage{mathrsfs}
\usepackage{stackengine}

\usepackage{float}
\usepackage{flushend, cuted}
\usepackage{lipsum}
\usepackage{lingmacros}
\usepackage{tree-dvips}
\usepackage{varioref}
 \usepackage{colortbl}
\usepackage{tikz}
\usetikzlibrary{shapes,arrows}
\usepackage{verbatim}
\usepackage{tikz}
\usetikzlibrary{shapes,arrows}
\usepackage{graphicx}
\usepackage{amsfonts}
\usepackage{amssymb}
\usepackage{lettrine}
 \usepackage{newlfont}
 \usepackage{mathrsfs}
 \usepackage{cite}
 \usepackage{booktabs}
 \usepackage{mathabx}
 \usepackage{multirow}%\usepackage{makeidx}
 \usepackage{nomencl}%   nomenclature generation via makeindex
 \usepackage{fancyhdr}
\usepackage{arydshln}
\usepackage{tabularx}
 \usepackage{colortbl}
\usepackage{graphicx}

\usepackage{bm}
\usepackage{caption}
\begin{document}
\title{Observer-Based Discontinuous Communication in the Secondary Control of AC Microgrids}
\author{ Shahabeddin Najafi, Yazdan Batmani, Pouya Shafiee, and Charalambos Konstantinou
\vspace{-8mm}
%\thanks{Shahabeddin Najafi and Yazdan Batmani are with Smart/Micro Grids Research Center (SMGRC), Department of Electrical Engineering, University of Kurdistan, Sanandaj, Iran. Pouya Shafiee is a Ph.D student at Department of Electrical Engineering, Virginia Tech, Blacksburg, VA, USA (e-mail: shahabedin.najafi@uok.ac.ir; y.batmani@uok.ac.ir; pouyashafiee@vt.edu).
}
\maketitle
\cfoot{\small 979-8-3503-9042-1/24/\$31.00 ©2024 IEEE}
\begin{abstract}
 This paper proposes an observer-based event-driven approach to decrease the overuse of communication networks. The suggested approach aims to estimate the required data for sharing between units in line with as much communication reduction as possible. In other words, the proposed approach effectively determines which state variables should be shared (observer concept) among the units during specific time intervals (event-triggered concept). This strategy significantly reduces the overall communication load. It is shown that the estimation error remains bounded and Zeno behavior, characterized by an endless number of transmissions occurring within a limited time frame, does not occur. The proposed methodology can be systematically applied to any communication-based secondary controller in alternating current (AC) microgrids. Simulation results demonstrate a high degree of precision in estimating the states under the proposed approach. Also, the secondary controller performance under the proposed method is evaluated in MATLAB/Simulink environment.
\end{abstract}
\vspace{-1mm}
\section{Introduction}\label{Sec1}
\IEEEPARstart{I}{n} Alternating current (AC) microgrids, acceptable voltage and frequency control, precise power-sharing, and effective reactive power filtering are key challenges that need to be addressed \cite{uzair2023challenges}. Different control structures have been employed to address these concerns in which the hierarchical control is recognized as a comprehensive architecture for control of microgrids \cite{abhishek2020review}. This structure in microgrids consists of four levels: primary, secondary, central/emergency, and global in which the last two structures are related to economic operation and interaction management among interconnected microgrids, respectively. The primary control is employed for voltage and frequency regulation to maintain primary stability but can cause deviations. Different architectures of the secondary control, including centralized, decentralized, and distributed, are considered to compensate for these deviations \cite{gao2019primary}. %It should be mentioned that the distributed one is widely employed due to its superior performance compared to its counterparts \cite{khayat2020decentralized}.

In communication-based secondary controllers, data exchange traditionally relies on a periodic method called ``time-triggered" communication with constant sampling times \cite{obermaisser2018event}. However, the sampling rate is required to be high enough to have the smallest error between the sampled data and their actual values. This type of communication results in the waste of energy by exchanging unnecessary information \cite{ge2020dynamic}.

In networked control systems, event-triggered communication is employed to overcome the time-triggered communication problems \cite{batmani2017event,bai2023dynamic,shafiee2021design,batmani2022improved}. This method reduces communication by transmitting data only when a defined error goes beyond a certain limit, avoiding unnecessary data transmission \cite{batmani2021event}. Event-triggered techniques have recently been applied in microgrids, serving as networked control systems \cite{fan2016distributed,weng2018distributed,qian2019event,choi2021distributed,ma2021accurate,doostinia2022distributed}. The paper \cite{fan2016distributed} utilizes nonlinear state-feedback based on an event strategy to enable the effective sharing of reactive power among the components.
In reference \cite{weng2018distributed}, two types of event-based control strategies are employed to restore the deviated frequency and voltage, as well as achieve precise power-sharing in AC microgrids. Reference \cite{qian2019event} uses two centralized and distributed event-triggered controllers for voltage and frequency control. Authors in \cite{choi2021distributed} consider a finite-time event-driven control to attain objectives in AC microgrids.
In \cite{ma2021accurate}, an event-driven leader-following consensus algorithm is designed to mitigate voltage deviation.  

The primary contribution of this research is the introduction of a fundamental event-based system for the secondary control of AC microgrids, regardless of the specific method used. The proposed event-based approach utilizes a full-state Luenberger observer to reduce pressure over the communication network. An important advantage of this strategy is its applicability to all forms of communication-based secondary control. Therefore, in comparison with other works, this method is not a secondary control method but is a proposed mechanism to provide the preliminary materials for any type of secondary controller in AC microgrids to conserve the network energy resources and deal with communication challenges by reducing usage of the communication network.
%The proposed event-triggered scheme uses a Luenberger observer placed by the $\mathrm{DG}_j$ to estimate all the states of the $\mathrm{DG}_i$ using the received data, which corresponds to the output of $\mathrm{DG}_i$ at triggering moments. Moreover, $\mathrm{DG}_i$ utilizes an event mechanism to determine the instances that its measured output should be transferred to the $\mathrm{DG}_j$. 
%It shows that the estimation error is bounded and avoids Zeno behavior. The method is tested in an AC microgrid using MATLAB/Simulink to validate its theoretical merits.}

Section \ref{Sec2} presents the mathematical model of distributed generators (DGs). Section \ref{Sec3} introduces the suggested event-based secondary control of voltage. Section \ref{Sec4} shows simulation results of implementing the proposed technique in an AC microgrid, and Section \ref{Sec5} concludes the paper.
%\vspace{-3mm}
\begin{figure*}[!ht]
\centering
\captionsetup{justification=centering}
\includegraphics[width=0.8\textwidth, trim=1.5px 1.5px 1.5px 1.5px, clip=true]{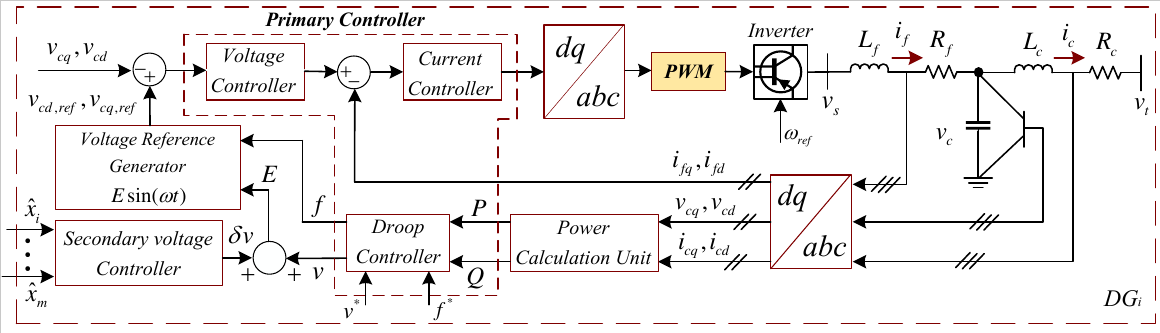}
\caption{The general control diagram of  AC microgrids for a DG.\vspace{-3mm}}\label{f1}
\end{figure*}
\section{Mathematical Modeling of DGs}\label{Sec2}
In this section, we derive a mathematical model for a DG based on the circuit diagram in Fig. \ref{f1}. By defining the state vector $x(t)={[i_\mathrm{fd}(t) \ \ i_\mathrm{fq}(t) \ \ v_\mathrm{cd}(t) \ \ v_\mathrm{cq}(t) \ \ i_\mathrm{cd}(t) \ \ i_\mathrm{cq}(t)]}^\mathrm{T}$, the input signal $u(t)={[v_\mathrm{sd}(t) \ \ v_\mathrm{sq}(t)]}^\mathrm{T}$, and the system output $y(t)=v_\mathrm{cd}(t)$, the system can be represented in the following state-space form:
\begin{eqnarray}\label{e2}
  \begin{split}
\dot{x}(t)&=\textbf{A}x(t)+\textbf{B}u(t)+\textbf{d}v_\mathrm{t_{dq}}(t),\\
y(t)&=\textbf{C}x(t),
\end{split}
\end{eqnarray}
where $v_\mathrm{s}(t)$, $v_\mathrm{c}(t)$, and $v_\mathrm{t}(t)$ are the inverter output voltage, the filter capacitor voltage, and the DG output voltage, respectively; $i_\mathrm{f}(t)$ and $i_\mathrm{c}(t)$ are the filter current and the output current, respectively. The indices $d$ and $q$ show the elements of the $dq$ coordinate. $R_\mathrm{f}$, $L_\mathrm{f}$, $C_\mathrm{f}$, $R_\mathrm{f}$, and $L_c$ are the filter resistance, the filter inductance, the filter capacitance, the line resistance, and the inductance, respectively. In this context, $\omega$ states the angular frequency in radians per second. Also, $\textbf{A}$, $\textbf{B}$, $\textbf{d}$, and $\textbf{C}$ are as follows:
\begin{align*}
\textbf{A}&=\!\!\begin{bmatrix}
\dfrac{-R_\mathrm{f}}{L_\mathrm{f}} & \omega & \dfrac{-1}{L_\mathrm{f}} & 0 & 0 & 0\\-\omega & \dfrac{-R_\mathrm{f}}{L_\mathrm{f}} & 0 & \dfrac{-1}{L_\mathrm{f}} & 0 & 0\\ \dfrac{1}{C_\mathrm{f}} & 0 & 0 & \omega & \dfrac{-1}{C_\mathrm{f}} & 0\\ 0 & \dfrac{1}{C_\mathrm{f}} & -\omega & 0 & 0 & \dfrac{-1}{C_\mathrm{f}}\\ 0 & 0 & \dfrac{1}{L_\mathrm{c}} & 0 & \dfrac{-R_\mathrm{c}}{L_\mathrm{c}} & \omega\\ 0 & 0 & 0 & \dfrac{1}{L_\mathrm{c}} & -\omega & \dfrac{-R_\mathrm{c}}{L_\mathrm{c}}
\end{bmatrix}\!,\:
 \textbf{C}\!\!=\!\!{\begin{bmatrix}
0 \\ 0 \\ 1 \\ 0 \\ 0 \\ 0
\end{bmatrix}}^\mathrm{T}\\[2pt]
{\textbf{d}}&= \left[ {\begin{array}{*{20}{c}}
{{0_{4 \times 1}}}&{{0_{4 \times 1}}}\\
{ - L_c^{ - 1}}&0\\
0&{ - L_c^{ - 1}}
\end{array}} \right],\:\:\: {\textbf{B}} = \left[ {\begin{array}{*{20}{c}}
{L_f^{ - 1}}&0\\
0&{L_f^{ - 1}}\\
{{0_{4 \times 1}}}&{{0_{4 \times 1}}}
\end{array}} \right].
\end{align*}

\vspace{2mm}
\section{Proposed Method}\label{Sec3}
 This paper aims to address the following questions in order to achieve significant communication reduction while maintaining performance:
\begin{itemize}
\item \textbf{$Q_1$}: Is it possible for each DG not to send all of its state variables? If so, which states should be transmitted through the communication network? (observer duty)\vspace{+1mm}
\item \textbf{$Q_2$}: Is it possible to send these states just at some specific time instants? (event-trigger mechanism duty)
\end{itemize}
%Without loss of generality, we begin by examining an AC microgrid consisting of two DGs. Subsequently, we expand the proposed event-triggered mechanism to accommodate AC microgrids with any number of DGs. Based on the model \eqref{e2}, it can be seen that each DG have six state variables. By answering \textbf{$Q_1$}, the $\mathrm{DG}_1$ might only send some of its states to the second DG and vice versa. Indeed, by designing a proper observer for $\mathrm{DG}_1$ ($\mathrm{DG}_2$), it is possible to estimate all its state variables and use them by $\mathrm{DG}_2$ ($\mathrm{DG}_1$). Therefore, the responsibility of observer is accomplished. The time sequence $\{t_k\}\ \ k\! \in\! \mathbb{Z}_{\geq0}$ should be designed so that $n^\prime$ states ($n^\prime<6$ is the number of the states which must be sent from $\mathrm{DG}_1$ to $\mathrm{DG}_2$) are transferred only at time instant $t_k$. This is the answer of the second question above under the event-triggered responsibility. In closing, by answering \textbf{$Q_1$} and \textbf{$Q_2$}, the size of the transmitted packets and its number are managed, respectively. 
%\begin{figure}[b!]
%\vspace{-5mm}
%\centering
%\includegraphics[width=3.5in, trim=1.5px 2.5px 1.5px 1.5px, clip=true]{1}
%\caption{General scheme of an AC microgrid}\label{f2}
%\end{figure}
%Now, consider an AC microgrid with $N$ DGs, as depicted in Fig. \ref{f2} under the secondary controller with architecture ``all-to-all" averaging communication in which all DGs are connected to each other in order to exchange their states.

\begin{figure}[h!]
\vspace{-3mm}
\centering
\captionsetup{justification=centering}
\includegraphics[width=2.6in, trim=1.0px 2.5px 1.5px 1.5px, clip=true]{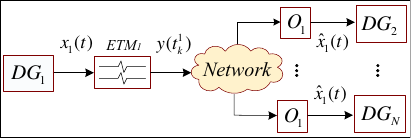}
\caption{The frame of the suggested event-based strategy.\vspace{-5mm}}\label{f3}
\end{figure}

Fig. \ref{f3} shows an event-triggered mechanism with $\mathrm{ETM}_1$ placed by $\mathrm{DG}_1$ to compute its triggering times $t_k^1,\ \ k\! \in\! \mathbb{Z}_{\geq0}$. $x_1(t)$ and $\hat{x}_1(t)$ represent the state vector of $\mathrm{DG}_1$ and its corresponding estimation, respectively; $y_1(t_k^1)$ is the transmitted output of $\mathrm{DG}_1$ to other agents. Furthermore, an observer denoted as $\mathrm{O}_1$ is specifically formed to estimate the states of $\mathrm{DG}_1$ based on the information provided by $y_1(t_k^1)$. A similar observer is implemented for the other units.
%\vspace{-1mm}
\subsection{Observer design}\label{subsec1}

Considering the following system:
\begin{eqnarray}
\label{e3}
  \begin{split}
\dot{x}(t)&=\textbf{A}x(t)+\textbf{B}u(t)\\
y(t)&=\textbf{C}x(t),
\end{split}
\end{eqnarray}
the state, control input, and measured output vectors are represented by $x(t)$, $u(t)$, and $y(t)$, respectively. $\textbf{A}$, $\textbf{B}$, $\textbf{C}$ are three constant matrices as well. The objective of this section is to design the following observer that can estimate the system state using $y(t)$ \cite{chen1999linear}:
\begin{eqnarray}\label{eq6}
  \begin{split}
\dot{\hat{x}}(t)&=\textbf{A}\hat x(t)+\textbf{B}u(t)+\textbf{L}(y(t)-\textbf{C}\hat{x}(t)),
\end{split}
\end{eqnarray}
which $\hat x(t)$ represents the estimated state vector, and $\textbf{L}$ denotes the observer gain matrix. By appropriately designing $\textbf{L}$ such that eigenvalues of $\textbf{A} - \textbf{LC}$ have negative real parts, the error $e(t) = x(t) - \hat{x}(t)$ tends to zero as $t\rightarrow \infty$. The arbitrary selection of these eigenvalues is determined by the choice of $\textbf{L}$ if $(\textbf{A},\textbf{C})$ is observable \cite{chen1999linear}.
%To check the state observability of $(\textbf{A},\textbf{C})$, consider the following matrix.
%\begin{equation*}
%\bm{\Phi}_o=\begin{bmatrix}
%\textbf{C}^{T} & (\textbf{CA})^{T} & \ldots & (\textbf{CA}^{n-1})^{T}
%\end{bmatrix}^{T}.
%\end{equation*}
%The pair $(\textbf{A},\textbf{C})$ is state observable if and only if $\bm{\Phi}_o$ is full rank . Assume that $(\textbf{A},\textbf{C})$ is state observable. 
To determine $\textbf{L}$, it is necessary to find a unique solution for $\textbf{S}=\textbf{S}^{T}>0$ using the following algebraic equation  \cite{batmani2021event}:
\begin{eqnarray}\label{eq7}
  \begin{split}
\textbf{AS}+\textbf{SA}^{T}-\textbf{SC}^{T}\textbf{V}^{-1}\textbf{CS}+\textbf{W}=\textbf{0},
\end{split}
\end{eqnarray}
where $\textbf{V}>0$ and $\textbf{W}\geq0$ are user-defined matrices. Thus, it can be demonstrated that $\textbf{A}-\textbf{LC}$ is a Hurwitz matrix when $\textbf{L}=\textbf{SC}^{T}\textbf{V}^{-1}$. It is feasible to design a stable Luenberger observer and estimate all the $i^\mathrm{th}$ DG states if we only have a $\textbf{C}_i$ matrix in a way that the pair ($\textbf{A}_i,\textbf{C}_i$) is fully state observable (in this case, $\textbf{C}_i=\begin{bmatrix}
0&0&1&0&0&0 \end{bmatrix}$ which measures the voltage $v_\mathrm{cd}(t)$).
%Lemma 1 is used to answer the first question ($Q_1$) at the beginning of this section. If ($\textbf{A}_i,\textbf{B}_i,\textbf{C}_i$) is the state space representation of the $i^\mathrm{th}$ DG, we should only select the measured matrix $\textbf{C}_i$ such that the pair ($\textbf{A}_i,\textbf{C}_i$) is fully state observable. Using \eqref{e1}, it can be checked that $\textbf{C}_i=\begin{bmatrix}
%0&0&1&0&0&0 \end{bmatrix}$ fulfills this observability condition. 
\vspace{-4mm}
\subsection{Event-based observer}
In accordance with \eqref{eq6}, if the output data is transmitted to the observer at $t_{k}$, $k \in \mathbb{Z}_{\geq0}$, this equation is stated as
\begin{eqnarray}\label{eq8}
  \begin{split}
\dot{\hat{x}}(t)&=\textbf{A}\hat x(t)+\textbf{B}u(t)+\textbf{LC}(x(t_{k})-\hat x(t)).
\end{split}
\end{eqnarray}
for $t \in [t_{k},t_{k+1})$. The sequence of $t_k$ should be obtained to guarantee the asymptotic stability of the system. To this end, consider $V(e(t))\!=\!e^{T}(t)\textbf{P}e(t)$ as the Lyapunov function candidate \cite{batmani2021event}. Its time derivative is stated as
\begin{eqnarray}
\label{eq11}
  \begin{split}
\dot{V}(e(t))&=\dot{e}^{T}(t)\textbf{P}e(t)+e^{T}(t)\textbf{P}\dot{e}(t).
\end{split}
\end{eqnarray}
Regarding the above-mentioned observer error term, once the event mechanism is not used, we have
\begin{eqnarray}\label{eq12}
\dot{e}(t)=\textbf{A}x(t)-\textbf{A}\hat{x}(t)-\textbf{LC}(x(t)-\hat{x}(t)).
\end{eqnarray}
%\par By defining $\textbf{A}_{c}\!=\!\textbf{A}-\textbf{LC}$ as well as adding and subtracting $\textbf{LC}x(t)$ to the right hand side of \eqref{eq12}, this relation can be written as follows:
By defining $\textbf{A}_{c}\!=\!\textbf{A}-\textbf{LC}$, \eqref{eq12} can be written as 
\begin{eqnarray}\label{eq13}
%\dot{e}(t)=\textbf{A}_{c}e(t)+\textbf{LC}\tilde{e}(t).
\dot{e}(t)=\textbf{A}_{c}e(t).
\end{eqnarray}
Substituting \eqref{eq13} in \eqref{eq11} results in 
\begin{align}\label{eq14}
%\dot{V}(e(t))=&{e}^{T}(t)(\textbf{A}_{c}^{T}\textbf{P}+\textbf{PA}_{c})e(t)+e^{T}(t)\textbf{PLC}\tilde{e}(t) \nonumber\\
%+&\tilde{e}^{T}(t)(\textbf{PLC})^{T}e(t).
\dot{V}(e(t))=&{e}^{T}(t)(\textbf{A}_{c}^{T}\textbf{P}+\textbf{PA}_{c})e(t) 
\end{align}
Let $\tilde{\textbf{Q}}$ be a positive-definite matrix, and the symmetric matrix $\textbf{P}$ presents the unique positive-definite solution of the Lyapunov equation given by:
\begin{equation}\label{eq10}
(\textbf{A}-\textbf{LC})^{T}\textbf{P}+\textbf{P}(\textbf{A}-\textbf{LC})=-\tilde{\textbf{Q}}.
\end{equation}
According to \eqref{eq10}, \eqref{eq14} can be written as follows for $t\geq 0$:
\begin{eqnarray}\label{eq15}
\dot{V}(e(t_k))=-{e}^{T}(t)\tilde{\textbf{Q}}e(t).
\end{eqnarray}
To tolerant a slower rate of decrease in $V(e(t))$, the following inequality should be hold \cite{batmani2021event}:
\begin{eqnarray}\label{eq16}
\dot{V}(e(t))\leq -\sigma{e}^{T}(t)\tilde{\textbf{Q}}e(t),
\end{eqnarray}
where $0<\sigma<1$ is a constant. When the event mechanism is used, \eqref{eq12}, \eqref{eq13}, and \eqref{eq14} can be rewritten as follows respectively.
\begin{align}\label{eq122}
\dot{e}(t)&=\textbf{A}x(t)-\textbf{A}\hat{x}(t)-\textbf{LC}(x(t_k)-\hat{x}(t)),\nonumber\\
\dot{e}(t)&=\textbf{A}_{c}e(t)+\textbf{LC}\tilde{e}(t),\nonumber\\
\dot{V}(e(t))&={e}^{T}(t)(\textbf{A}_{c}^{T}\textbf{P}+\textbf{PA}_{c})e(t)+e^{T}(t)\textbf{PLC}\tilde{e}(t) \nonumber\\
&+\tilde{e}^{T}(t)(\textbf{PLC})^{T}e(t). 
\end{align}
where $\tilde{e}(t)=x(t)-x(t_{k})$. By replacing \eqref{eq122} in \eqref{eq16}, we have
\begin{eqnarray}\label{eq17}
(\sigma-1){e}^{T}(t)\tilde{\textbf{Q}}e(t)+2{e}^{T}(t)\textbf{PLC}\tilde{e}(t)\leq 0,
\end{eqnarray}
which can be rewritten as 
\begin{eqnarray}\label{eq9}
  \begin{split}
\eta (t) = &\begin{bmatrix}
e^{T}(t) & \tilde{e}^{T}(t)
\end{bmatrix}\bm{\Psi}\begin{bmatrix}
e(t) \\ \tilde{e}(t)
\end{bmatrix}\leq 0
\end{split}
\end{eqnarray}
where
\begin{equation*}
\bm{\Psi}=\begin{bmatrix}
(\sigma-1)\tilde{\textbf{Q}} & \textbf{PLC} \\ (\textbf{PLC})^{T} & \textbf{0}
\end{bmatrix}
\end{equation*}
Therefore, the time sequence $t_k$ can be obtained as
\begin{eqnarray}\label{eq100}
t_{k+1}={\inf}\{t>0\:|\: t>t_{k} \:\wedge\:\eta (t)= 0\}.
\end{eqnarray} 
%Consider the represented linear system by \eqref{e2} in which the pair $(\textbf{A},\textbf{C})$ is state observable and the observer gain $\textbf{L}=\textbf{SC}^{T}\textbf{V}^{-1}$, $\textbf{S}>0$ is the unique solution of \eqref{eq7} \cite{batmani2021event}.
%Then $t_{k}$, $k\in \mathbb{Z}_{\geq0}$ is obtained once the following event condition is violated:
\begin{eqnarray}\label{eq9}
  \begin{split}
\eta (t) = &\begin{bmatrix}
e^{T}(t) & \tilde{e}^{T}(t)
\end{bmatrix}\bm{\Psi}\begin{bmatrix}
e(t) \\ \tilde{e}(t)
\end{bmatrix}\leq 0
\end{split}
\end{eqnarray}

To experience further communication reduction, an additional condition can be considered for the event-detector mechanism. In this regard, the variable $t_k$ is modified as \cite{batmani2021event}
\begin{eqnarray}\label{eq100a}
  \begin{split}
t_{k+1}={\inf}\{t>0\:|\: t>t_{k} \:\wedge\:\eta (t)= 0\:\:\&\:\: e(t) \notin B_\epsilon\},
\end{split}
\end{eqnarray}
where
\begin{eqnarray}\label{eq110a}
  \begin{split}
B_\epsilon = \{e(t) \in \mathbb{R}^n\:|\: \vert\vert e(t)\vert\vert \leq \epsilon \},
\end{split}
\end{eqnarray}
and $\epsilon$ is a user-defined positive constant. It is demonstrated that $e(t)$ is globally uniformly ultimately bounded (UUB) by satisfying the event condition specified in \eqref{eq100a}, as discussed in \cite{batmani2021event}. Furthermore, the Zeno-free characteristic of the proposed event mechanism has been mathematically proved \cite{batmani2021event}.
%\begin{table}[t!]
%\caption{\textcolor{blue}{Values of the simulation parameters.}}
%\label{table1}
%\centering
%\includegraphics[width=2.6in, trim= 3px 3px 3px 3px, clip=true]{kkkkk.png}
%\end{table}
\vspace{-1mm}
\subsection{Proposed secondary control of voltage}
Using the droop mechanism, the relation between reactive power and voltage can be represented by
\begin{equation}\label{eq30}
v(t)=v^*(t)-n_\mathrm{d}Q(t)+\delta v(t)
\end{equation}
where \!$v(t)$, $v^*(t)$, $n_\mathrm{d}$, and \!$Q(t)$ are the voltage, the reference voltage, the voltage droop coefficient, and the reactive power, respectively. Using the secondary control, the correction term $\delta v(t)$ is produced to eliminate the voltage deviation induced by the droop controller.  In \eqref{eq30}, $Q(t)$ takes the following form:
\begin{equation}\label{eq31}
Q(t)=v_\mathrm{cq}(t)i_\mathrm{cd}(t)-v_\mathrm{cd}(t)i_\mathrm{cq}(t).
\end{equation}

\par
\begin{figure}[b!]
\vspace{-2mm}
\centering
\captionsetup{justification=centering}
\includegraphics[width=2.3in, trim=1.5px 1.5px 1.5px 3.5px, clip=true]{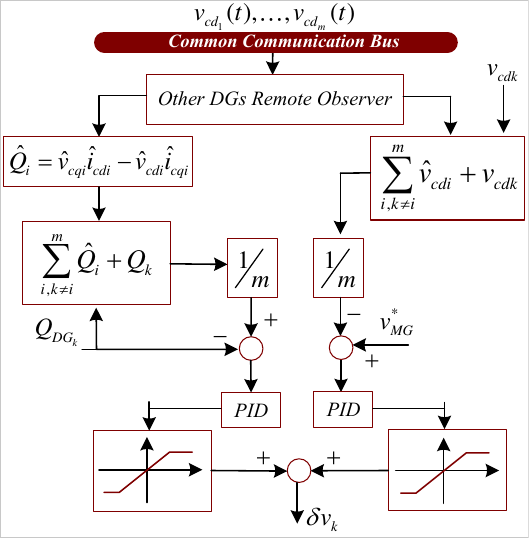}
 \caption{Event-triggered observer-based secondary controller.\vspace{-5mm}}\label{f5}
\end{figure}

 In the secondary control with all-to-all averaging communication, each DG requires accesses other DGs' data to produce $\delta v(t)$. %Consider the ``all-to-all" averaging communication, which is shown in Fig. \ref{f4}. 
%In this structure, the secondary controller of each DG collects the measured voltage amplitudes and reactive power of the other DGs through the communication network, average them, and produces the proper control signal ($\delta v(t)$) for the primary level.% As mentioned at the end of Subsection \ref{subsec1}, if $(A_i,C_i)$ is state observable, it is possible to use the proposed observer-based event-triggered mechanism. 
Fig. \ref{f5} indicates an observer-based all-to-all secondary controller equipped by an event-triggered structure. In this figure, $\delta v_l$ is acquired by means of the estimated states, i.e., $\hat{v}_{\mathrm{cd}_i}$ and $\hat{Q}_i=\hat{v}_{\mathrm{cq}_i}\hat{i}_{\mathrm{cd}_i}-\hat{v}_{\mathrm{cd}_i}\hat{i}_{\mathrm{cq}_i}$, $i=1,...,N$. The proposed method enables the 
 transmission of only one state ($v_\mathrm{cd}$) from a DG to its neighbors. This state is only sent during event instants $t_k$, reducing the usage of the communication network.
\begin{figure}[b!]
\vspace{-5mm}
\centering
\captionsetup{justification=centering}
\includegraphics[width=3.1in, trim= 3px 3px 3px 3px, clip=true]{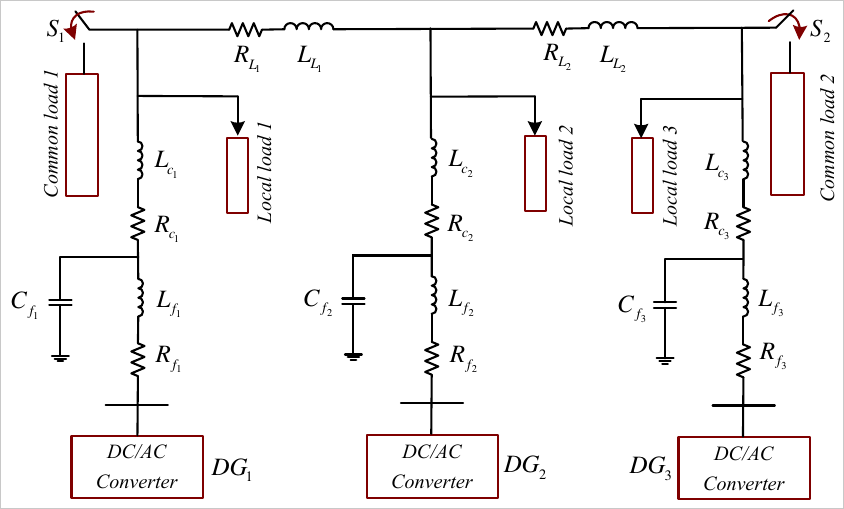}
 \caption{Diagram of the considered AC microgrid.}\label{f6}
\end{figure}
%\begin{figure}[t!]
%\centering
%\includegraphics[width =3.55in, trim=1.5px 1.5px 1.5px 1.5px, clip=true]{5}
 %\caption{Diagram of the ``all-to-all" averaging secondary controller of voltage.}\label{f4}
%\end{figure}

\section{Simulation Results}\label{Sec4}
This section evaluates the proposed method in a $311 \:\mathrm{V}$ AC microgrid with three DGs (Fig. \ref{f6}). 
The frequency of the microgrid is assumed to be $50\:\mathrm{Hz}$ as well. Design weighting matrices $W$ and $V$ in the observer are considered as $10^7I_2$ and $I_2$, respectively.
%. Also, $L_\mathrm{c}=L_\mathrm{f}=1.8 \: \mathrm{mH}$,  $R_\mathrm{c}=R_\mathrm{f}=0.1 \: \mathrm{\Omega}$,  $C_\mathrm{f}=3 \: \mathrm{\mu F}$, $R_\mathrm{L_1}=0.23 \ \mathrm{\Omega}$, $L_\mathrm{L_1}=318.3 \ \mathrm{mH}$, $R_\mathrm{L_2}=0.35 \ \mathrm{\Omega}$, $L_\mathrm{L_2}=1.8 \ \mathrm{mH}$. Local loads are $LL_1=8 \ \mathrm{kW},\ 6.9 \ \mathrm{kVAr}$, $LL_2=0 \  \mathrm{kW},\ 0 \  \mathrm{kVAr}$, and $LL_3=7.5\ \mathrm{kW},\ 6 \  \mathrm{kVAr}$. Also, common loads are $CL_1=5\ \ \mathrm{kW}, \ 4 \ \mathrm{kVAr}$, and $CL_2=4 \ \mathrm{kW}, \ 3 \ \mathrm{kVAr}$. 
Active and reactive droop coefficients are $m_\mathrm{d}=9.4\times10^{-5}$ and $n_\mathrm{d}=1.3\times10^{-3}$, respectively. 
In voltage proportional–integral–derivative (PID) controller, proportional and internal terms are $k_\mathrm{pV}=1$ and $k_\mathrm{iV}=5$. Also, in reactive PID controller, proportional and integral terms are $k_\mathrm{pQ}=0.003$ and $k_\mathrm{iQ}=0.4$, respectively. 
The required simulation parameters are listed in Table \ref{table1}. The results are presented in the following two subsections. In Subsections \ref{subsec11} and \ref{subsec12}, the state estimation and the secondary control of voltage based on estimated states are evaluated, respectively.
\begin{table}[b!]
\vspace{-1mm}
\captionsetup{justification=centering}
\caption{Values of the simulation parameters.}
\label{table1}
\vspace{-3mm}
\setlength{\tabcolsep}{2mm}
\renewcommand\arraystretch{0.9}
\begin{center}
\begin{tabular}{|c|c|c|}
\hline
\textbf{Symbol} & \textbf{Description} & \textbf{Value} \\ \hline\hline
$R_\mathrm{f}$ & Filter resistance & $0.1 \ \ \mathrm{\Omega}$ \\ \cline{1-3}
$L_\mathrm{f}$ & Filter inductance & $1.8 \ \ \mathrm{mH}$ \\ \cline{1-3} 
$C_\mathrm{f}$ & Filter capacitance & $3 \ \ \mathrm{\mu F}$ \\ \cline{1-3} 
$R_\mathrm{c}$ & Output resistance & $0.1 \ \ \mathrm{\Omega}$ \\ \cline{1-3} 
$L_\mathrm{c}$ & Output inductance & $1.8 \ \ \mathrm{mH}$ \\ \cline{1-3} 
$R_\mathrm{L_1}$ & Line 1 resistance & $0.23 \ \ \mathrm{\Omega}$ \\ \cline{1-3} 
$L_\mathrm{L_1}$ & Line 1 inductance & $318.3 \ \ \mathrm{mH}$ \\ \cline{1-3} 
$R_\mathrm{L_2}$ & Line 2 resistance & $0.35 \ \ \mathrm{\Omega}$ \\ \cline{1-3} 
$L_\mathrm{L_2}$ & Line 2 inductance & $1.8 \ \ \mathrm{mH}$ \\ \cline{1-3} 
$LL_1$ & Local load 1 & \begin{tabular}[c]{@{}c@{}}$8 \ \ \mathrm{kW}$\\ $6.9 \ \ \mathrm{kVAr}$\end{tabular} \\ \cline{1-3} 
$LL_2$ & Local load 2 & \begin{tabular}[c]{@{}c@{}}$0 \ \ \mathrm{kW}$\\ $0 \ \ \mathrm{kVAr}$\end{tabular} \\ \cline{1-3}    
$LL_3$ & Local load 3 & \begin{tabular}[c]{@{}c@{}}$7.5 \ \ \mathrm{kW}$\\ $6 \ \ \mathrm{kVAr}$\end{tabular} \\ \cline{1-3} 
$CL_1$ & Common load 1 & \begin{tabular}[c]{@{}c@{}}$5 \ \ \mathrm{kW}$\\ $4 \ \ \mathrm{kVAr}$\end{tabular} \\ \cline{1-3} 
$CL_2$ & Common load 2 & \begin{tabular}[c]{@{}c@{}}$4 \ \ \mathrm{kW}$\\ $3 \ \ \mathrm{kVAr}$\end{tabular} \\ \hline
%$m_\mathrm{d}$ & Active power droop coefficient & \begin{tabular}[c]{@{}c@{}}$9.4\times10^{-5}$\\ $\mathrm{Wsec/rad}$\end{tabular} \\[5pt] \cline{1-3} 
%$n_\mathrm{d}$ & Reactive power droop coefficient & \begin{tabular}[c]{@{}c@{}}$1.3\times10^{-3}$\\ $\mathrm{VAr/V}$\end{tabular} \\
%$k_\mathrm{pV}$ & Voltage PID proportional term & $1$ \\[5pt] \cline{1-3} 
%$k_\mathrm{iV}$ & Voltage PID integral term & $5 \ \ \mathrm{sec}^{-1}$ \\[5pt] \cline{1-3} 
%$k_\mathrm{pQ}$ & \begin{tabular}[c]{@{}c@{}}Reactive power PID\\ proportional term\end{tabular} & $0.003$ \\[5pt] \cline{1-3} 
%$k_\mathrm{iQ}$ & Reactive power PID integral term & $0.4 \ \ \mathrm{sec}^{-1}$ \\[5pt] \hline
%$W$ & Design weighting matrix & $10^{7}I_{2}$ \\[8pt] \cline{1-3} 
%$V$ & Design weighting matrix & $I_{2}$ \\ \hline
\end{tabular}
\end{center}
\end{table}
\vspace{-5mm}
\subsection{Evaluating of the simulated state estimation}\label{subsec11}
\begin{comment}
\begin{figure*}[t!]
\centering
\includegraphics[width=\textwidth, trim=1.5px 1.5px 1.5px 1.5px, clip=true]{d1}
 \caption{States of the DGs and their estimations.}\label{f8}
 \vspace{-3mm}
\end{figure*}
\end{comment}

\begin{figure}[t!]
\centering
%\vspace{-5mm}
\includegraphics[width=3.1in, trim=1.5px 1.5px 1.5px 1.5px, clip=true]{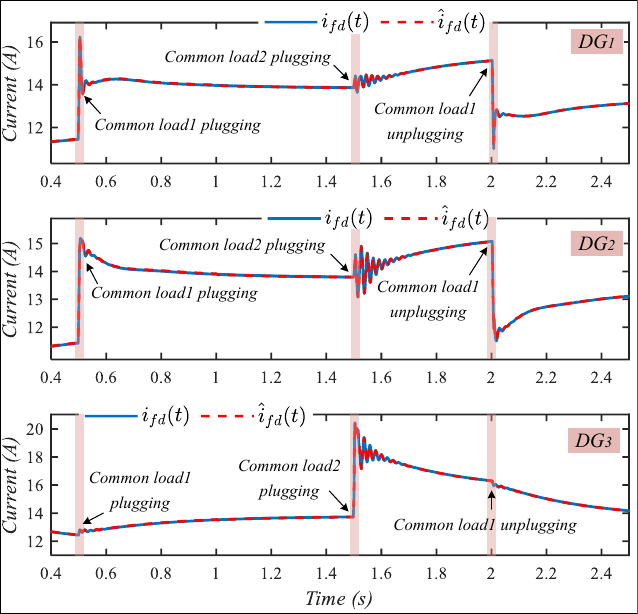}
 \vspace{-1mm}
\captionsetup{justification=centering}
 \caption{DGs' filter current and their estimations in $d$ frame.}\label{f8}
 \vspace{-3mm}
\end{figure}

\begin{figure}[t!]
\centering
\includegraphics[width=3.1in, trim=1.5px 1.5px 1.5px 1.5px, clip=true]{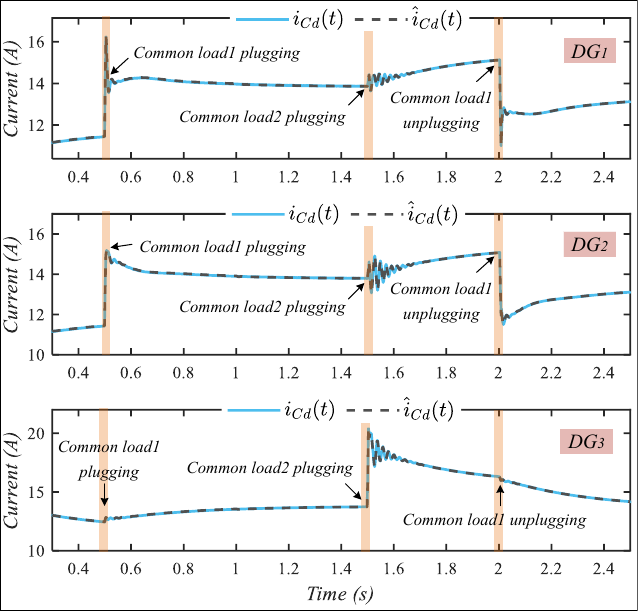}
\vspace{-1mm}
\captionsetup{justification=centering}
 \caption{DGs' output currents and their estimations in $d$ frame.\vspace{-3mm}}\label{f81}
 \vspace{-2mm}
\end{figure}

\begin{figure}[t!]
\centering
\includegraphics[width=3in, trim=1.5px 1.5px 1.5px 1.5px, clip=true]{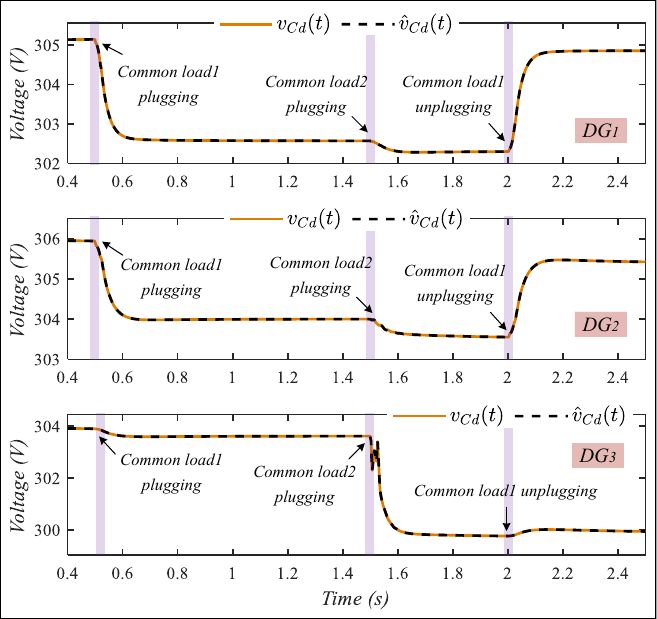}
\vspace{-1mm}
\captionsetup{justification=centering}
 \caption{DGs' output voltage and their estimations in $d$ frame.}\label{f82}
 \vspace{-2mm}
\end{figure}

\begin{figure}[h!]
\centering
\includegraphics[width=3in, trim=1.5px 1.5px 1.5px 1.5px, clip=true]{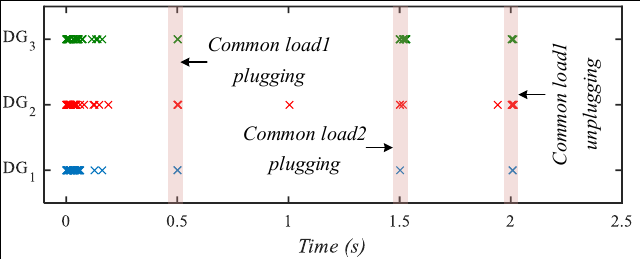}
\vspace{-1mm}
\captionsetup{justification=centering}
 \caption{Diagrams of event instants for all DGs.}\label{f14}
 \vspace{-2mm}
\end{figure}

In this subsection, there are two different scenarios. In the first scenario, all DGs initially supply their local loads without any common loads ($S_1$ and $S_2$ are open). At $t=0.5\ \mathrm{s}$, $S_1$ is closed. At $t=1.5\ \mathrm{s}$, $S_2$ is also closed and the second common load is plugged into the grid. Finally, at $t=2\ \mathrm{s}$, the first common load is disconnected from the grid. Figs. \ref{f8}, \ref{f81}, and \ref{f82} show the diagrams of states and their estimations for each DG in $d$ frame in which all estimated states converge to their actual values with acceptable accuracy. Fig. \ref{f14} shows the capability of the proposed method in a considerable reduction of communication. As anticipated, the rate of communication is considerably reduced during periods without changes in the common loads. Data exchange only occurs during the switching instances, demonstrating the efficient communication strategy of the proposed method.
\vspace{-5mm}
\begin{figure}[h!]
\centering
\includegraphics[width=3in, trim=1.5px 1.5px 1.5px 1.5px, clip=true]{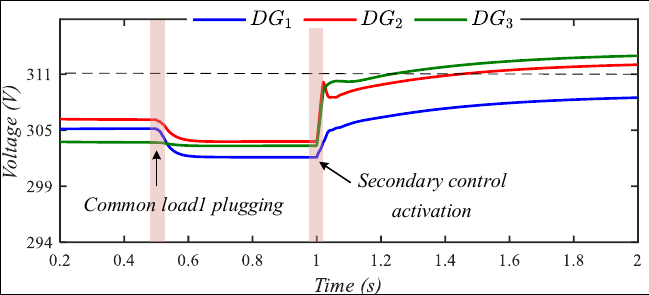}
\vspace{-1mm}
\captionsetup{justification=centering}
 \caption{Output voltage in the presence of proposed controller.\vspace{-5mm}}\label{f17}
 %\vspace{-1mm}
\end{figure}
\begin{figure}[t!]
\centering
\includegraphics[width=3in, trim=1.5px 1.5px 1.5px 1.5px, clip=true]{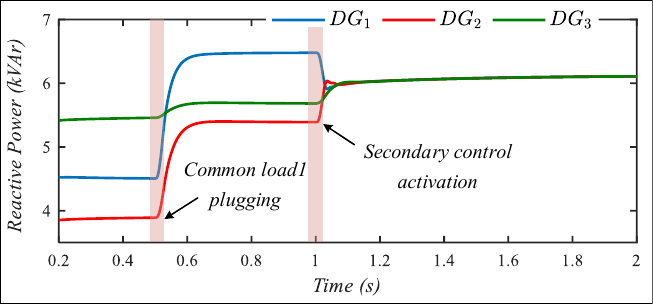}
%\vspace{-1mm}
\captionsetup{justification=centering}
 \caption{DGs' reactive power with the proposed controller.\vspace{-5mm}}\label{f18}
 %\vspace{-2mm}
\end{figure}
\vspace{4mm}
\subsection{Secondary control evaluation}\label{subsec12}
The proposed observer-based event-driven secondary controller is applied to the same AC microgrid in the previous subsection. Initially, two switches $S_1$ and $S_2$, are open. Then, at $t\!=\!0.5\ \mathrm{s}$, $S_1$ is closed, followed by activating the proposed secondary controller of voltage at $t\!=\!1\ \mathrm{s}$. Fig. \ref{f17} illustrates the diagram of $v_{cd}(t)$ for all units, where the controller effectively compensates for the voltage deviations. Fig. \ref{f18} displays the diagrams of the reactive power for all the DGs as well. The proposed secondary control successfully restores voltage and ensures proper sharing of reactive power among the DGs.
%\vspace{-3mm}
\section{Conclusion}\label{Sec5}
The proposed observer-based event-triggered scheme offers an efficient solution for voltage restoration and reactive power-sharing in AC microgrids. By implementing an event-triggered strategy, data exchange only occurs during switching instances. The embedded observer decreases the number and size of shared packets within the microgrid by preventing units from sending all their data during event times, resulting in a further reduction in communication. The estimation error remains bounded, and the occurrence of Zeno behavior is avoided. Simulation results demonstrate the effectiveness of the proposed strategy in significantly reducing network usage while maintaining microgrid stability and performance. %Considering the communication challenges such as packets dropout, time latency, noise, etc., in presence of the proposed method and how it can cope withe them is an extension of this work.
This methodology, being a universal strategy applicable to various secondary controllers, does not primarily focus on the challenges of secondary control. Hence, an extension of this work would be to examine its ability to manage communication issues such as packet loss, time delay, and measurement noise.
\vspace{-5mm}
\bibliographystyle{ieeetr}
\bibliography{NewRef}
\end{document}